\documentclass[12pt]{article}
\usepackage{amssymb}

\begin{document}

\title{\textbf{Lifting Frenet Formulas}}
\author{Mehmet Tekkoyun}
\date{}
\maketitle
\date{}

\begin{abstract}
In this study, we conclude the vertical, complete and horizontal lifts of
Frenet formulas given by (1) and defined on space $\mathbf{R}^{3}$ to its
tangent space $T\mathbf{R}^{3}=\mathbf{R}^{6}$.

\textbf{M.S.C. 2000:} 57R25, 28A51

\textbf{Key words:} vector fields, frame fields, lift theory
\end{abstract}

\textbf{1. Introduction}

Let a unit speed curve $\beta _{0}(t)$ with constant $\kappa >0$ on space $%
\mathbf{R}^{3}$, and suppose that $T,N,B$ be respectively tangent, normal,
binormal vector on any point of $\beta _{0}(t).$ Then, we will call that the
triple $\{T,N,B\}$ is \textit{Frenet frame} such that $T.N=B.T=B.N=0$. In
differentiable geometry, lift method plays an important role. Because, it is
possible to generalize to differentiable structures on any space (resp.
manifold) to extended spaces (resp. extended manifolds) using lift function
\cite{yano,tekkoyun1,tekkoyun2}. So, it may be extended the following
theorem given on space $\mathbf{R}^{3}$ to its tangent space $T\mathbf{R}%
^{3} $.

\textit{Theorem 1.1(The Frenet Formulas)}\cite{opera}\textit{. }For a unit
speed curve $\beta _{0}(t)$ with constant $\kappa >0$ on $\mathbf{R}^{3}$,
the derivatives of Frenet frame $\{T,N,B\}$ are given by

\begin{equation}
T^{^{\prime }}=\kappa N,\,\,N^{\prime }=-\kappa T+\tau _{0}B,\,\,B^{\prime
}=-\tau _{0}N  \label{1.1}
\end{equation}

where $T,N,B,\kappa ,\tau _{0}$ is respectively tangent vector, normal
vector, binormal vector, curvature, torsion of the curve $\beta _{0}(t).$

The paper is structured as follows. In second 2, the vertical, complete and
horizontal lifts of a vector field defined on any manifold $M$ of dimension $%
m$ and their lift properties will be extended to space $T\mathbf{R}^{3}$. In
second 3, vertical lift of the theorem above will be obtained. Then, similar
to vertical lift, complete and horizontal lift analogues of the related
theorem are given.

In this study, all geometric objects will be assumed to be of class $%
C^{\infty }$ and the sum is taken over repeated indices. Also, $v,c$, and $H$
denote the vertical, complete and horizontal lifts of any differentiable
geometric structures defined on $\mathbf{R}^{3}$ to $T\mathbf{R}^{3},$
respectively.

\textbf{2}. \textbf{Lift of Vector Field}

The \textit{vertical lift }of a vector field $X$ on space $\mathbf{R}^{3}$
to extended space $T\mathbf{R}^{3}(=\mathbf{R}^{6})$ is the vector field $%
X^{v}\in \chi (T\mathbf{R}^{3})$ given as:
\begin{equation}
X^{v}(f^{c})=(Xf)^{v},\forall f\in \digamma (\mathbf{R}^{3})  \label{2.1}
\end{equation}

The vector field $X^{c}\in \chi (T\mathbf{R}^{3})$ defined by
\begin{equation}
X^{c}(f^{c})=(Xf)^{c},\forall f\in \digamma (\mathbf{R}^{3})  \label{2.2}
\end{equation}

is called the \textit{complete lift }of a vector field $X$ on $\mathbf{R}%
^{3} $ to its tangent space $T\mathbf{R}^{3}$.

The \textit{horizontal lift }of a vector field $X$ on a space $\mathbf{R}%
^{3} $ to $T\mathbf{R}^{3}$ is the vector field $X^{H}\in \chi (T\mathbf{R}%
^{3})$ determined by
\begin{equation}
X^{H}(f^{v})=(Xf)^{v},\forall f\in \digamma (\mathbf{R}^{3}).  \label{2.3}
\end{equation}

The general properties of vertical, complete and horizontal lifts of a
vector field on $\mathbf{R}^{3}$ are as follows:

\textbf{Proposition 2.1:} Let be functions all\thinspace \thinspace
\thinspace $f,g\in \digamma (\mathbf{R}^{3})\,\,$and vector fields all $%
X,Y\in \chi (\mathbf{R}^{3}).\,\,$Then it is satisfied the following
equalities.

\begin{equation}
\begin{array}{ll}
i) & \,\,(X+Y)^{v}=X^{v}+Y^{v},(X+Y)^{c}=X^{c}+Y^{c},(X+Y)^{H}=X^{H}+Y^{H},
\\
ii) & \,\,\,(fX)^{v}=f^{v}X^{v},(fX)^{c}=f^{c}X^{v}+f^{v}X^{c},(fg)^{H}=0,
\\
iii) & \,\,\,X^{v}(f^{v})=0,%
\,X^{c}(f^{v})=X^{v}(f^{c})=(Xf)^{v},X^{c}(f^{c})=(Xf)^{c},%
\,X^{H}(f^{v})=(Yf)^{v}, \\
v) & \;\chi (U)=Sp\left\{ \frac {\partial }{\partial x^{\alpha }}\right\}
,\,\chi (TU)=Sp\left\{ \frac{\partial }{\partial x^{\alpha }},\frac{\partial
}{\partial y^{\alpha }}\right\} , \\
& \,\,(\frac{\partial }{\partial x^{\alpha }})^{c}=\frac{\partial }{\partial
x^{\alpha }},(\frac{\partial }{\partial x^{\alpha }})^{v}=\frac{\partial }{%
\partial y^{\alpha }},(\frac{\partial }{\partial x^{\alpha }})^{H}=\frac{%
\partial }{\partial x^{\alpha }}-\Gamma _{\beta }^{\alpha }\frac{\partial }{%
\partial y^{\alpha }}.%
\end{array}
\label{2.4}
\end{equation}

Where $\Gamma _{\beta }^{\alpha }$ are Christopher symbols, $U$ and $TU$\
are respectively topological opens of $\mathbf{R}^{3}$ and $T\mathbf{R}%
^{3},\,$ $f^{v},f^{c}\in \digamma (T\mathbf{R}^{3}),%
\,X^{v},Y^{v},X^{c},Y^{c},X^{H},Y^{H}\in \chi (T\mathbf{R}^{3}),$ $1\leq
\alpha ,\beta \leq 3.$

\textbf{3.} \textbf{Lifting Frenet Formulas}

In this section, we compute the vertical, complete and horizontal lifts of
Frenet formulas given by means of $T,N$ and $B$ Frenet vectors on a unit
speed curve $\beta _{0}(t)$ with constant $\kappa >0$ defined on space $%
\mathbf{R}^{3}$.

\textbf{a)} \textbf{The vertical lifting Frenet Formulas}

Let $T^{v}$ be vertical lift of tangent vector $T$ on a unit speed curve $%
\beta _{0}(t).$ Length of $T^{v}$ is given as:
\[
\left| \left| T^{v}\right| \right| =T^{v}T^{v}=(T.T)^{v}=1
\]

With respect to the product rule, it follows

\[
0=(T^{v}T^{v})^{^{\prime }}=(T^{v})^{^{\prime
}}.T^{v}+T^{v}.(T^{v})^{^{\prime }}=2T^{v}.(T^{v})^{^{\prime }}.
\]

Thus $T^{v}.(T^{v})^{^{\prime }}=0$ and $(T^{v})^{^{\prime }}$ is found
orthonormal to $T^{v}.$ Therefore it is said $(T^{v})^{^{\prime }}$ is
normal to unit speed curve $\beta _{1}(t)=(\beta _{0}(t))^{v}.$ Similarly,
we have

\[
B^{v}.T^{v}=B^{v}.N^{v}=0.
\]

In this case $T^{v},N^{v}$ and $B^{v}$ are three orthonormal Frenet vectors
on $\beta _{1}(t)$ in the 6-dimensional space $T\mathbf{R}^{3}.$

\textit{Theorem 3.1: }For a unit speed curve $\beta _{1}(t)$ with constant $%
(\kappa )^{v}>0$ on $T\mathbf{R}^{3}$, the derivatives's vertical lifts of
the Frenet vectors are given as:
\begin{equation}
(T^{\prime })^{v}=(\kappa )^{v}N^{v},\,\,(N^{\prime })^{v}=-(\kappa
)^{v}T^{v}+(\tau _{0})^{v}B^{v},\,\,(B^{\prime })^{v}=-(\tau _{0})^{v}N^{v}
\label{3.1}
\end{equation}

Where $(\kappa )^{v}$ $=\left| \left| (T^{^{\prime }})^{v}\right| \right| $
and $(\tau _{0})^{v}=-N^{v}.(B^{\prime })^{v}$ are respectively curvature
and torsion of the curve $\beta _{1}(t).$

\textit{Proof: }Let\textit{\ }$(T^{^{\prime }})^{v},(N^{^{\prime
}})^{v},(B^{\prime })^{v}$ be vertical lifts of $T^{^{\prime }},N^{^{\prime
}}$, $B^{^{\prime }}$ which are derivatives of $T,N,B,$ respectively. We
already know
\begin{equation}
(T^{^{\prime }})^{v}=(\kappa )^{v}N^{v}  \label{3.2}
\end{equation}
by definition of $N^{v}$, where the curvature $(\kappa )^{v}$describes
variation in direction of $T^{v}$. Also, we shall find $(B^{\prime })^{v}$
and $(N^{^{\prime }})^{v}.$ In particular, Given $(B^{\prime
})^{v}=a_{1}T^{v}+b_{1}N^{v}+c_{1}B^{v}.$ If it can be identified $%
a_{1},b_{1},c_{1},T^{v},N^{v}$ and $B^{v}$ then it will be known $(B^{\prime
})^{v}.$ Firstly, we have
\begin{eqnarray}
T^{v}.(B^{\prime })^{v} &=&a_{1}T^{v}.T^{v}+b_{1}T^{v}.N^{v}+c_{1}T^{v}.B^{v}
\label{3.3} \\
&&a_{1}(T.T)^{v}+b_{1}(T.N)^{v}+c_{1}(T.B)^{v}  \nonumber \\
&=&a_{1}.1+b_{1}.0+c_{1}.0  \nonumber \\
&=&a_{1}.  \nonumber
\end{eqnarray}

Similarly, $N^{v}.(B^{\prime })^{v}=b_{1}$ and $B^{v}$.$(B^{\prime
})^{v}=c_{1}.$ So, it follows
\begin{equation}
(B^{\prime })^{v}=(T^{v}.(B^{\prime })^{v})T^{v}+(N^{v}.(B^{\prime
})^{v})N^{v}+(B^{v}.(B^{\prime })^{v})B^{v}.  \label{3.4}
\end{equation}

Now let's identify $T^{v}.(B^{\prime })^{v}.$ We know $%
T^{v}.B^{v}=(T.B)^{v}=0,$ so that $0=(T^{v}B^{v})^{\prime }=(T^{^{\prime
}})^{v}.B^{v}+T^{v}.(B^{\prime })^{v}$ by vertical lift properties and the
product rule. Then, using $N^{v}.B^{v}=(N.B)^{v}=0,$ it is found
\begin{equation}
T^{v}.(B^{\prime })^{v}=-(T^{^{\prime }})^{v}.B^{v}=-(\kappa
)^{v}N^{v}.B^{v}=0.  \label{3.5}
\end{equation}

We can also identify $B^{v}.(B^{\prime })^{v}.$ We have $%
B^{v}.B^{v}=(B.B)^{v}=1$, so $0=(B^{v}.B^{v})^{\prime }=(B^{\prime
})^{v}.B^{v}+B^{v}.(B^{\prime })^{v}=2B^{v}.(B^{\prime })^{v}.$ Thus, $%
(B.B^{\prime })^{v}=B^{v}.(B^{\prime })^{v}=0.$ Define $(\tau
_{0})^{v}=-N^{v}.(B^{\prime })^{v}$ be the torsion of the curve $\beta
_{1}(t).$ From the above, $(B^{\prime })^{v}$ is calculated as:

\begin{equation}
(B^{\prime })^{v}=-(\tau _{0})^{v}N^{v}  \label{3.6}
\end{equation}

Now it will be obtained $(N^{\prime })^{v}.$ Just as for $(B^{\prime })^{v},$
it follows
\begin{equation}
(N^{\prime })^{v}=(T^{v}.(N^{\prime })^{v})T^{v}+(N^{v}.(N^{\prime
})^{v})N^{v}+(B^{v}.(N^{\prime })^{v})B^{v}.  \label{3.7}
\end{equation}

The same types of calculations give $(T.N)^{v}=T^{v}.N^{v}=0$, therefore $%
0=(T^{\prime })^{v}.N^{v}+T^{v}.(N^{\prime })^{v}$ and $(T^{\prime
})^{v}=(\kappa )^{v}N^{v}$ so it is obtained $T^{v}.(N^{\prime
})^{v}=-(\kappa )^{v}N^{v}.N^{v}=-(\kappa )^{v}.$ Also, $N^{v}.N^{v}=1$, so $%
N^{v}.(N^{\prime })^{v}=0$ and $B^{v}.N^{v}=0$, in this case $(B^{\prime
})^{v}.N^{v}+B^{v}.(N^{\prime })^{v}=0.$ Thus, by definition it is found to
be $B^{v}.(N^{\prime })^{v}=-(B^{\prime })^{v}.N^{v}=-N^{v}.(B^{\prime
})^{v}=(\tau _{0})^{v}$. Hence, $(N^{^{\prime }})^{v}$ is computed to be
\begin{equation}
(N^{^{\prime }})^{v}=-(\kappa )^{v}T^{v}+(\tau _{0})^{v}B^{v}.  \label{3.8}
\end{equation}

Therefore, proof finishes.$\Box $

\textbf{b) The complete and horizontal lifting Frenet formulas}

One may easily show that it is the complete and horizontal lift analogues of
Frenet formulas the following as:

\textit{Theorem 3.2: }For a unit speed curve $\beta _{2}(t)=(\beta
_{0}(t))^{c}$ with constant $(\kappa )^{c}>0$ on tangent space $T\mathbf{R}%
^{3}$, complete lifts of the derivatives of the Frenet frame are given by
equations

\begin{equation}
(T^{^{\prime }})^{c}=(\kappa )^{c}N^{c},\,\,(N^{^{\prime }})^{c}=-(\kappa
)^{c}T^{c}+(\tau _{0})^{c}B^{c},\,\,(B^{^{\prime }})^{c}=-(\tau
_{0})^{c}N^{c}  \label{3.9}
\end{equation}

where $(\kappa )^{c}$ $=\left| \left| (T^{^{\prime }})^{c}\right| \right| $%
and $(\tau _{0})^{c}=-N^{c}.(B^{^{\prime }})^{c}$ are curvature and torsion
of the curve $\beta _{2}(t),$ respectively$.$

\textit{Theorem 3.3: }For a unit speed curve $\beta _{3}(t)=(\beta
_{0}(t))^{H}$ with $(\kappa )^{H}>0$ on the tangent space $T\mathbf{R}^{3}$,
the expression's horizontal lifts of derivatives of the Frenet frame are
equalized expressions
\begin{equation}
(T^{^{\prime }})^{H}=(\kappa )^{H}N^{H},\,\,(N^{^{\prime }})^{H}=-(\kappa
)^{H}T^{H}+(\tau _{0})^{H}B^{H},\,\,(B^{^{\prime }})^{H}=-(\tau
_{0})^{H}N^{H},  \label{3.10}
\end{equation}

where the curvature and torsion of the curve $\beta _{3}(t)\,\,$are
respectively given by $(\kappa )^{H}$ $=\left| \left| (T^{^{\prime
}})^{H}\right| \right| $and $(\tau _{0})^{H}=-N^{H}.(B^{^{\prime }})^{H}$ $%
.\,$

\textbf{Conclusion}

In this study, using lifting methods, we see that it may be generalized the
derivatives of Frenet frame elements given a unit speed curve on space $%
\mathbf{R}^{3}$ to its extension $T\mathbf{R}^{3}.$

Author's address:

Mehmet Tekkoyun

Pamukkale University, Faculty of Science \& Art, Mathematics Department,
Kinikli Campus, 20070 Denizli, Turkey.

E-mail: tekkoyun@pamukkale.edu.tr\thinspace \thinspace \thinspace \thinspace
\thinspace \thinspace \thinspace \thinspace \thinspace

\end{document}